\documentclass[letterpaper,preprint,aps]{revtex4-1}
\usepackage[latin9]{inputenc}
\usepackage{amsmath}
\usepackage{amssymb}

\makeatletter

\pdfpageheight\paperheight
\pdfpagewidth\paperwidth

\usepackage{graphicx}
\usepackage{tikz}
\usepackage{bbm}
\usepackage{verbatim}
\usepackage[percent]{overpic}

\DeclareMathOperator*{\diag}{diag}

\DeclareMathOperator{\sech}{sech}

\def\dd{{\rm d}}

\makeatother

\begin{document}
\title{Conformal Wavefunctions for Graviton Amplitudes}
\author{Chang Liu}
\email{chang\_liu3@brown.edu}

\author{David A. Lowe}
\email{lowe@brown.edu}

\affiliation{Department of Physics, Brown University, Providence, RI, 02912, USA}
\begin{abstract}
The extended-BMS algebra of asymptotically flat spacetime contains
an $SO(3,1)$ subgroup that acts by conformal transformations on the
celestial sphere. It is of interest to study the representations of
this subgroup associated with gravitons. To reduce the equation of
motion to a Schrodinger-like equation it is necessary to impose a
non-covariant gauge condition. Using these solutions, leading-order
gauge invariant Weyl scalars are then computed and decomposed into
families of unitary principal series representations. An invertible
holographic mapping is constructed between these unitary principal
series operators and massless spin-2 perturbations of flat spacetime.
\end{abstract}
\maketitle

\section{Introduction}

In previous work \cite{Liu:2021tif}, massive scalar fields in 4D
Minkowski spacetime were decomposed into modes on 3D de-Sitter spacetime
slices where they form unitary principal series representations of
${\rm SO(3,1)}$. This study was motivated by the program of \cite{Pasterski:2016qvg}
where the goal is to formulate gravity in asymptotically flat spacetime
as a theory on the celestial sphere with conformal symmetry. In this
paper we extend this construction to massless spin-2 particles, or
gravitons, in 4D Minkowski spacetime. To this end, we consider linearized
gravitational waves living in flat 4D background spacetime with the
standard spherical coordinates. The background metric is simply 
\[
g_{\mu\nu}=\diag(-1,1,r^{2},r^{2}\sin^{2}\theta)
\]
Following the notations in \cite{Bernar:2014lna}, from now on indices
$a,b,c,\ldots$ refer to the ``orbit'' spacetime labeled by the
coordinates $(t,r)$, and $i,j,k,\ldots$ refer to the 2-sphere labeled
by $(\theta,\varphi)$. In other words, we write the background metric
as 
\[
g_{\mu\nu}\,\dd x^{\mu}\dd x^{\nu}=g_{ab}\,\dd y^{a}\dd y^{b}+r^{2}\,\dd\sigma^{2}
\]
where $g_{ab}=\diag(-1,1)$ and $\dd\sigma^{2}=\gamma_{ij}\,\dd z^{i}\dd z^{j}=\dd\theta^{2}+\sin^{2}\theta\,\dd\varphi^{2}$.

For Minkowski spacetime in four dimensions, the gravitational perturbations
can be expanded in terms of both the scalar and the vector spherical
harmonics defined on the 2-sphere. These are also known as the ``even''
and ``odd'' waves in \cite{PhysRev.108.1063}, and will automatically
have the desired transformation properties under the rotation group
${\rm SO(3)}$. However, as we shall see, the specific gauge conditions
that we will choose in what follows are not Lorentz covariant, and
therefore these metric perturbations do not transform as tensors under
the full ${\rm SO(3,1)}$ group. To remedy this, we consider the Newman-Penrose
formalism \cite{doi:10.1063/1.1724257} of general relativity and
construct leading-order gauge-invariant scalars known as the Weyl
scalars. These scalars can then be mapped onto the celestial sphere
through a generalization of the method described in the previous work
\cite{Liu:2021tif}. This generalization involves performing a spectral
decomposition into radial eigenvalues using the Meijer K-transform
\cite{meijer1940a,meijer1940b}. We therefore obtain an invertible
holographic map between graviton fields on the flat Minkowski background
spacetime and conformal operators on the celestial sphere. We stress
that this procedure is defined for tree-level amplitudes and it remains
to be seen whether an interacting holographic theory can be defined
independently of the four-dimensional gravitational description.

\section{Graviton Wavefunctions}

\subsection{Scalar perturbations}

Refs.~\cite{Kodama:2003jz,doi:10.1142/S0218271816410169} present
a general formalism which expresses the metric perturbation $h_{\mu\nu}$
in terms of a master function $\phi$. For the scalar perturbation,
a gauge choice allows us to express the metric perturbation $h_{\mu\nu}$,
expanded in terms of the spherical harmonics $Y_{lm}$, as 
\begin{equation}
h_{ab}=f_{ab}Y_{lm},\quad h_{ai}=0,\quad h_{ij}=f\,\gamma_{ij}Y_{lm}\label{eq:ansatz}
\end{equation}
where $f_{ab}$ and $f$ are functions that are related to a master
function $\phi(t,r)$ 
\begin{align}
f & =\frac{l(l+1)r\phi}{2}+r^{2}\partial_{r}\phi\nonumber \\
f_{ab} & =\partial_{a}\partial_{b}(r\phi)-\frac{g_{ab}}{2}\Box(r\phi)\nonumber \\
 & =r\partial_{a}\partial_{b}\phi+\partial_{a}r\,\partial_{b}\phi+\partial_{b}r\,\partial_{a}\phi-\frac{g_{ab}}{2}(r\Box\phi+2\partial_{r}\phi)\,.\label{eq:fab}
\end{align}
Here $\partial_{a}r\partial_{b}\phi$ stands for $(\partial_{a}r)(\partial_{b}\phi)$
and $\Box$ is the d'Alembertian on the orbit spacetime. The components
of $f_{ab}$ are 
\begin{align*}
f_{tt} & =f_{rr}=r\partial_{t}^{2}\phi+\partial_{r}\phi+\frac{l(l+1)}{2r}\phi\\
f_{tr} & =r\partial_{t}\partial_{r}\phi+\partial_{t}\phi\,.
\end{align*}
The master function $\phi(t,r)$ satisfies the following master equation
\[
\Box\phi-\frac{l(l+1)}{r^{2}}\phi=0
\]
which can be solved to yield a basis of mode functions 
\[
\phi_{\omega l}(t,r)=e^{-i\omega t}\sqrt{r}(c_{1}J_{l+\frac{1}{2}}(\omega r)+c_{2}Y_{l+\frac{1}{2}}(\omega r))
\]
for $\omega\neq0$. Here $J_{n}$ is the Bessel function of first
kind and $Y_{n}$ is the Bessel function of second kind. We demand
that the mode functions be regular at the origin, and therefore discard
the second set of solutions. We therefore find the (unnormalized)
modes for the master function 
\[
\phi_{\omega l}(t,r)=e^{-i\omega t}\sqrt{r}J_{l+\frac{1}{2}}(\omega r)\,.
\]
For $\omega\neq0$ this agrees with the master equation found in Ref.~\cite{PhysRevLett.24.737}.

For $\omega=0$ we have the special time-independent solution of the
master equation 
\[
\phi_{0l}(t,r)=c_{1}r^{l+1}+c_{2}r^{-l}\,.
\]
Discarding solutions that are divergent at $r\to\infty$ we find the
basis of functions for the $\omega=0$ modes 
\[
\phi_{0l}(t,r)=r^{-l}\,.
\]
Physically this represents a time-independent spacetime perturbation
that is rotating at a constant angular momentum (for $l\neq0$. For
$l=0$ the metric perturbation is zero), similar to the eternal Kerr
black hole which appears to a distant observer as having a total angular
momentum. Substituting the master function into eq.~\ref{eq:ansatz}
we find the scalar metric perturbation for $\omega=0$ 
\[
h_{\mu\nu}=Y_{lm}\begin{bmatrix}\frac{l(l-1)}{2}r^{-l-1} & 0 & 0 & 0\\
0 & \frac{l(l-1)}{2}r^{-l-1} & 0 & 0\\
0 & 0 & \frac{l(l-1)}{2}r^{1-l} & 0\\
0 & 0 & 0 & \frac{l(l-1)}{2}r^{1-l}\sin^{2}\theta
\end{bmatrix}\,.
\]

\subsection{Vector perturbations}

We define the vector spherical harmonics as a vector field $Y_{i}^{(lm)}$
on the unit 2-sphere satisfying 
\[
[\Delta_{2}+(l(l+1)-1)]Y_{i}^{(lm)}=0
\]
with $D^{i}Y^{(lm)}{}_{i}=0$. Here $\Delta_{2}$ and $D_{i}$ are
the Laplace operator and the covariant derivative on the unit 2-sphere,
respectively. In terms of the scalar spherical harmonics $Y_{lm}$
we find 
\[
Y_{i}^{(lm)}(\theta,\varphi)=\frac{\epsilon_{ij}}{\sqrt{l(l+1)}}\partial^{j}Y_{lm}(\theta,\varphi)
\]
Here indices are raised and lowered using the metric $\gamma_{ij}$
on the unit 2-sphere, and $\epsilon^{ij}$ is the Levi-Civita tensor
on the 2-sphere defined by $\epsilon_{\theta\varphi}=\sqrt{|\det\gamma_{ij}|}=\sin\theta$.
Given suitable gauge choice \cite{PhysRev.108.1063}, the metric perturbation
can then be expanded in terms of $Y_{i}^{(lm)}$ as follows 
\begin{equation}
h_{ab}=0,\quad h_{ai}=f_{a}Y_{i}^{(lm)},\quad h_{ij}=0\label{eq:vectoransatz}
\end{equation}
where $f^{a}$ is related to the master function $\phi(t,r)$ above
via 
\[
f^{a}=\epsilon^{ab}\partial_{b}(r\phi)\,.
\]
Here $\epsilon_{ab}$ is the Levi-Civita tensor on the two-dimensional
orbit spacetime defined by $\epsilon^{tr}=+1$. For $\omega\neq0$
this agrees with the so-called odd waves in Ref.~\cite{PhysRev.108.1063}.
For $\omega=0$ substituting the master function $\phi=r^{-l}$ into
eq.~\ref{eq:vectoransatz} we find the metric perturbation 
\[
h_{\mu\nu}=\sqrt{\frac{l}{l+1}}r^{-l-1}\begin{bmatrix}0 & 0 & \frac{1}{\sin\theta}\partial_{\varphi}Y_{lm} & -\sin\theta\,\partial_{\theta}Y_{lm}\\
0 & 0 & 0 & 0\\
\frac{1}{\sin\theta}\partial_{\varphi}Y_{lm} & 0 & 0 & 0\\
-\sin\theta\,\partial_{\theta}Y_{lm} & 0 & 0 & 0
\end{bmatrix}\,.
\]

\section{Klein-Gordon inner product}

The metric perturbation expressions constructed above are yet to be
normalized. Following \cite{doi:10.1142/S0218271816410169} we define
the inner product between two metric perturbations $h_{\mu\nu}$ and
$h'_{\mu\nu}$ as 
\[
\langle h,h'\rangle=-i\int_{\Sigma}\dd\Sigma\,n_{\lambda}(h_{\mu\nu}^{\star}p'{}^{\lambda\mu\nu}-h'_{\mu\nu}p^{\star\lambda\mu\nu})
\]
where $\Sigma$ is a Cauchy surface and $n^{\lambda}$ is the future-directed
unit vector field normal to $\Sigma$. Here $p^{\lambda\mu\nu}$ is
the conjugate momentum current 
\[
p^{\lambda\mu\nu}=g^{\lambda\nu}\nabla_{\kappa}h^{\kappa\nu}+g^{\lambda\nu}\nabla_{\kappa}h^{\kappa\mu}-\nabla^{\lambda}h^{\mu\nu}+g^{\mu\nu}(\nabla^{\lambda}h-\nabla^{\kappa}h^{\lambda}{}_{\kappa})-\frac{g^{\lambda\nu}\nabla^{\mu}h+g^{\lambda\mu}\nabla^{\nu}h}{2}
\]
Here all indices are raised and lowered with respect the the background
metric $g_{\mu\nu}$ and $\nabla$ is the covariant derivative of
the background metric.

\subsection{Scalar perturbations}

For the scalar perturbation Ref.~\cite{doi:10.1142/S0218271816410169}
eq.~88 has shown that for $h_{\mu\nu}^{\omega lm}$ and $h_{\mu\nu}^{\omega'l'm'}$,
derived from the master function $\phi_{\omega l}$ and $\phi_{\omega'l'}$
via eq.~\ref{eq:ansatz} respectively, the conserved inner product
is 
\[
\langle h^{\omega lm},h^{\omega'l'm'}\rangle=-i\int_{0}^{+\infty}\dd r\,\delta_{ll'}\delta_{mm'}\,J^{0}
\]
where the orbit spacetime current $J^{a}$ is given by 
\[
J^{a}=\frac{4}{r}\partial^{c}r(f^{\star ab}f'_{bc}-f^{ab}f'{}_{bc}^{\star})-(f^{\star bc}\partial^{a}f'_{bc}-f'{}^{bc}\partial^{a}f_{bc}^{\star})\,
\]
Here $f_{ab}$ is related to the master function $\phi_{\omega l}$
and $f'_{ab}$ is related to the master function $\phi_{\omega'l'}$
via eq.~\ref{eq:fab}. Substituting eq.~\ref{eq:fab} and following
the derivations in eqs.~90--100 and Appendix B of Ref.~\cite{doi:10.1142/S0218271816410169}
we find 
\[
J^{0}=-\frac{l(l-1)(l+1)(l+2)}{2}(\phi_{\omega l}^{\star}\partial_{t}\phi_{\omega'l'}-\phi_{\omega'l'}\partial_{t}\phi_{\omega l}^{\star})\,.
\]
For $\phi_{\omega l}=e^{-i\omega t}\sqrt{r}J_{l+\frac{1}{2}}(\omega r)$
we have 
\[
i\,(\phi_{\omega l}^{\star}\partial_{t}\phi_{\omega'l}-\phi_{\omega'l}\partial_{t}\phi_{\omega l}^{\star})|_{t=0}=(\omega+\omega')rJ_{l+\frac{1}{2}}(r\omega)J_{l+\frac{1}{2}}(r\omega')\,.
\]
Using 
\[
\int_{0}^{\infty}rJ_{\alpha}(\omega r)J_{\alpha}(\omega'r)\,\dd r=\frac{\delta(w-w')}{\omega}
\]
we find 
\[
\langle h^{\omega lm},h^{\omega'l'm'}\rangle=l(l-1)(l+1)(l+2)\delta(\omega-\omega')\delta_{ll'}\delta_{mm'}\,.
\]
We see that in order to normalize the metric perturbations to have
\[
\langle h^{\omega lm},h^{\omega'l'm'}\rangle=\delta(\omega-\omega')\delta_{ll'}\delta_{mm'}
\]
we will perform a change of variable $h_{\mu\nu}\to[l(l-1)(l+1)(l+2)]^{-1/2}h_{\mu\nu}$.
In the interest of notational clarity we will assume this has been
done and will continue to use $h_{\mu\nu}$ to denote the normalized
scalar metric perturbation.

\subsection{Vector perturbations}

For the vector perturbations Ref.~\cite{doi:10.1142/S0218271816410169}
eq.~A10 has shown that for $h_{\mu\nu}^{V}$ and $\left.h'\right._{\mu\nu}^{V}$
given by master functions $\phi_{\omega lm}$ and $\phi_{\omega'l'm'}$
(via eq.~\ref{eq:vectoransatz}) we have 
\[
\langle h^{V},h'{}^{V}\rangle=-i\,\delta_{ll'}\delta_{mm'}(l-1)(l+2)\int_{0}^{+\infty}\dd r\,(\phi_{\omega l}^{\star}\partial_{t}\phi_{\omega'l'}-\phi_{\omega'l'}\partial_{t}\phi_{\omega l}^{\star})\,.
\]
Following the calculations of the previous section we find 
\[
\langle h^{V},\tilde{h}^{V}\rangle=2(l-1)(l+2)\delta(\omega-\omega')\delta_{ll'}\delta_{mm'}\,.
\]
We see that a change of variable $h_{\mu\nu}^{V}\to[2(l-1)(l+2)]^{-1/2}h_{\mu\nu}$
will allow us to normalize the vector metric perturbations to have
\[
\langle(h^{V})^{\omega lm},(h^{V})^{\omega'l'm'}\rangle=\delta(\omega-\omega')\delta_{ll'}\delta_{mm'}\,.
\]

\section{Gauge Invariant Observables}

The solutions we have described above depend on the choice of gauge
\eqref{eq:ansatz}. Since our graviton modes are expanded in terms
of scalar or vector spherical harmonics, we expect that they transform
under the rotation group ${\rm SO(3)}$ in the usual way. Indeed,
denoting $(h^{S})_{\mu\nu}^{\omega lm}$ by $|\omega lm\rangle$,
we find that 
\[
L_{3}\,|\omega lm\rangle=m\,|\omega lm\rangle
\]
and 
\[
(L_{1}\pm iL_{2})\,|\omega lm\rangle=\sqrt{(l\mp m)(l\pm m+1)}\,|\omega l,m\pm1\rangle
\]
where the rotation operators $L_{i}$ act on the graviton modes as
\[
L_{i}h_{\mu\nu}=-i\,[\epsilon_{ijk}x^{j}\partial^{k}h_{\mu\nu}+(\omega_{ij})_{\mu}{}^{\lambda}h_{\lambda\nu}+(\omega_{ij})_{\nu}{}^{\lambda}h_{\mu\lambda}]
\]
with 
\[
(\omega_{\mu\nu})_{\alpha}{}^{\beta}=\eta_{\mu\alpha}\delta_{\nu}^{\beta}-\eta_{\nu\alpha}\delta_{\mu}^{\beta}\,.
\]
However, since the gauge conditions (eqs.~\ref{eq:ansatz},\ref{eq:vectoransatz})
for either the scalar and vector perturbations are not Lorentz covariant,
we do not expect that these graviton modes transform as representations
of the full Lorentz group ${\rm SO(3,1)}$. Indeed, acting with the
boost operators 
\[
K_{i}h_{\mu\nu}=-i\,[(x^{0}\partial^{i}-x^{i}\partial^{0})h_{\mu\nu}+(\omega_{0i})_{\mu}{}^{\lambda}h_{\lambda\nu}+(\omega_{0i})_{\nu}{}^{\lambda}h_{\mu\lambda}]
\]
does not produce the correct Lorentz algebra. In order to restore
the correct Lorentz algebra, one needs to perform additional gauge
transformations after a boost to restore the gauge conditions (eqs.~\ref{eq:ansatz},\ref{eq:vectoransatz}).
The exact gauge transformations required are non-trivial and do not
have a closed expression as far as we know. 

Alternatively, one could consider the transverse-traceless gauge that
is indeed Lorentz covariant and rewrite our graviton modes in this
gauge, and attempt to quantize such a theory with a procedure similar
to the Gupta-Bleuler formalism of electromagnetism. However this procedure
does not provide a complete gauge fixing, and one is still left with
a Hilbert space containing zero-norm states. Neither of these approaches
will be completely satisfactory in producing a conformal description
of the graviton modes as gauge invariant operators on the celestial
sphere.

In the context of AdS/CFT the usual procedure would be to adopt Fefferman-Graham
coordinates where one can simply identify components of the metric
expansion around spatial infinity with a boundary stress-energy tensor.
The boundary stress-energy tensor then provides a complete description
of the boundary data for gravitational waves. 

Finding an analogous set of variables in the case of asymptotically
flat spacetime is a somewhat more thorny problem. One approach is
simply to choose a Bondi metric near null infinity \cite{doi:10.1098/rspa.1962.0161,doi:10.1098/rspa.1962.0206}
and describe the gravitational waves using the asymptotic variables
that appear there. Another approach, used in the numerical study of
gravitational waves from time-dependent collapsing/colliding objects,
is to instead pick a distinguished tetrad and compute the so-called
Weyl scalars (see for example \cite{PhysRevD.73.064005,Nerozzi:2016kky}.
An infinitesimal gauge transformation of the Riemann tensor is
\[
\delta R_{\mu\nu\beta}^{\alpha}=\mathcal{L}_{\xi}R_{\mu\nu\beta}^{\alpha}
\]
where $\xi$ parameterizes the diffeomorphism and $\mathcal{L}_{\xi}$
is the Lie derivative. Since the Lie derivative is linear in $R_{\mu\nu\beta}^{\alpha}$
and linear in $\xi$ this will vanish at leading order in $\xi$ if
the Riemann tensor is computed at linear order in the perturbation
around flat spacetime. So the Weyl scalars, which amount to picking
particular components of the Riemann tensor in this context, will
be a set of gauge invariant observables at leading order. 

We are therefore led to the consideration of the so-called spin-coefficient
formalism \cite{deFelice:1990hu} of general relativity, a special
example of which is known as the Newman-Penrose formalism \cite{doi:10.1063/1.1724257}.
Here one picks a null tetrad satisfying 
\[
l_{\mu}l^{\mu}=n_{\mu}n^{\mu}=m_{\mu}m^{\mu}=\bar{m}_{\mu}\bar{m}^{\mu}=0
\]
normalized so that
\[
l_{\mu}n^{\mu}=-1,\qquad m_{\mu}\bar{m}^{\mu}=1
\]
with other cross contractions between two vectors vanishing. For our
specific purpose we will pick the limit of the Kinnersley tetrad \cite{kinnersley}
\begin{align}
l^{\mu} & =\left(1,1,0,0\right)\,,\qquad n^{\mu}=\left(\frac{1}{2},-\frac{1}{2},0,0\right)\,,\nonumber \\
m^{\mu} & =\frac{1}{\sqrt{2}r}\left(0,0,1,\frac{i}{\sin\theta}\right)\,,\qquad\bar{m}^{\mu}=\frac{1}{\sqrt{2}r}\left(0,0,1,-\frac{i}{\sin\theta}\right)\,.\label{eq:kinner}
\end{align}
The five Weyl scalars $\Psi_{i}$ for $i=0,\ldots,4$ are built out
of the Weyl tensor $C_{\alpha\beta\gamma\delta}$ of the full spacetime
as 
\begin{align*}
\Psi_{0} & =C_{\alpha\beta\gamma\delta}\,l^{\alpha}m^{\beta}l^{\gamma}m^{\delta}\\
\Psi_{1} & =C_{\alpha\beta\gamma\delta}\,l^{\alpha}n^{\beta}l^{\gamma}m^{\delta}\\
\Psi_{2} & =C_{\alpha\beta\gamma\delta}\,l^{\alpha}m^{\beta}\bar{m}^{\gamma}n^{\delta}\\
\Psi_{3} & =C_{\alpha\beta\gamma\delta}\,l^{\alpha}n^{\beta}\bar{m}^{\gamma}n^{\delta}\\
\Psi_{4} & =C_{\alpha\beta\gamma\delta}\,n^{\alpha}\bar{m}^{\beta}n^{\gamma}m^{\delta}\,.
\end{align*}
Here we expand $C_{\alpha\beta\gamma\delta}$ to first order in $h_{\mu\nu}$.
Since the full spacetime satisfies the vacuum Einstein equation, we
find that the Weyl tensor $C_{\alpha\beta\gamma\delta}$ is equal
to the Riemann tensor $R_{\alpha\beta\gamma\delta}$. The Weyl scalars,
gauge invariant under infinitesimal gauge transformations, may then
be expressed in terms of the graviton wavefunctions of the previous
section. Our strategy will then be to decompose these gauge invariant
scalars into representations of the conformal group.

Note that for a particular value of $l,m$ it would be possible to
find modification of the tetrad, at leading order in the perturbation,
such as picking $l$ along a principal null direction, that would
make some of the Weyl scalars vanish. However if make pick the tetrad
independently of the perturbation, all the Weyl scalars will typically
be non-vanishing, and will provide a basis for gauge invariant observables
(at leading order).

A straightforward calculation leads to the expressions for the Weyl
scalars evaluated on the scalar and vector perturbations in appendices
A and B respectively. These expressions involve radial derivations
of the master function $\phi(t,r)$ and angular derivatives of the
spherical harmonics. For us, the main point is that the Weyl scalars
do not satisfy a simple wave equation in 4D spacetime. In order to
use the methods of \cite{Liu:2021tif} we next must perform a spectral
decomposition in the radial direction to allow us to use that basis
functions as a complete basis for the Weyl scalars.

\section{Holographic Mapping to the Celestial Sphere}

The Weyl scalars encode all the information of the gravitational perturbations.
Our strategy will be to proceed in two steps: first consider a fixed
radius 3D de Sitter slice of flat spacetime, and use the Plancherel
(or completeness) theorem for the unitary principle series representations
to decompose a general function on such a slice into irreducible representations;
next we allow for a general radial variation of such functions, effectively
decomposing a general solution into solutions of the 4D massive scalar
wave equations with a continuous spectrum of masses.

The starting point is the unitary principal series mode functions
that we have computed in \cite{Liu:2021tif}
\[
\Phi_{pMlm}(\eta,\rho,z,\bar{z})=\phi_{pl}(\eta)\,\psi_{pM}(\rho)\,Y_{m}^{l}(z,\bar{z})
\]
where 
\[
\phi_{pl}(\eta)=\sech\eta\left[\frac{i\pi}{2}P_{l}^{i\sqrt{p^{2}-1}}\left(\tanh\eta\right)+Q_{l}^{i\sqrt{p^{2}-1}}\left(\tanh\eta\right)\right]
\]
and 
\[
\psi_{pM}(\rho)=\frac{K_{i\sqrt{p^{2}-1}}(M\rho)}{\rho}\,.
\]
Here $M$ is the 4D scalar mass, $l,m$ are the usual angular momentum
quantum numbers, and $p$ labels the unitary principal series representation
and also behaves as a radial quantum number. $K_{\nu}(x)$ is the
modified Bessel function of second kind, and $(\eta,\rho,z,\bar{z})$
are the hyperbolic coordinates on Minkowski spacetime with metric
\[
\dd s^{2}=-\rho^{2}\dd\eta^{2}+\dd\rho^{2}+\rho^{2}\cosh^{2}\eta\,\frac{4\,\dd z\,\dd\bar{z}}{(1+|z|^{2})^{2}}\,.
\]
We will apply the main result of Ref.~\cite{Liu:2021tif}, which
states that these modes $\Phi_{pMlm}$ form a unitary principal series
representation of ${\rm SO(3,1)}$. This allows us to apply the Plancherel
theorem \cite{knapp2001representation} and map the Weyl scalars into
sets of conformal operators defined on the celestial sphere. We now
discuss this procedure in detail.

To motivate the full 4D map, let us first consider the simpler case
where one is to construct a holographic map on the 3D de-Sitter slice
(which we take to be the $\rho=1$ hypersurface) of 4D Minkowski.
On the 3D de-Sitter slice, the mode functions $\Phi_{pMlm}$ above
reduce to the mode functions $\phi_{plm}(\eta,z,\bar{z})=\phi_{pl}(\eta)Y_{lm}(z,\bar{z})$.
We can then use these modes as basis for the analogue of Fourier transform,
and maps a scalar function $f(\eta,z,\bar{z})$ into ``Fourier''
coefficients labeled by $\hat{f}(p,l,m)$ 
\[
\hat{f}(p,l,m)=\int_{-\infty}^{+\infty}\cosh^{2}\eta\,\dd\eta\int_{\mathbb{C}}\frac{4\dd z\dd\bar{z}}{(1+|z|^{2})^{2}}\,\bar{\phi}_{pl}(\eta)\,\bar{Y}_{lm}(z,\bar{z})\,f(\eta,z,\bar{z})\,.
\]
The Plancherel theorem \cite{knapp2001representation} guarantees
that this map is unitary. Using eq.~10.40 of Ref.~\cite{knapp2001representation},
we find the inverse map 
\[
f(\eta,z,\bar{z})=\int_{1}^{\infty}2p\tanh\left(\frac{\pi}{2}\sqrt{p^{2}-1}\right)\dd p\sum_{lm}\phi_{pl}(\eta)\,Y_{lm}(z,\bar{z})\,\hat{f}(p,l,m)\,.
\]
In particular, the integral measure of the inverse map follows from
setting $v=\sqrt{p^{2}-1}$ in $v\tanh(\pi v/2)\,\dd v$ of eq.~10.40
\cite{knapp2001representation}, and multiplying it by two, since
both $p\in(1,+\infty)$ and $p\in(-\infty,-1)$ map to $v\in(0,+\infty)$.

In order to include the radial variation of the scalars, we note the
fact that the following Meijer $K$-transform of order $\nu$, defined
for a function $f(x)$ as 
\[
\hat{f}(y)=\int_{0}^{\infty}f(x)K_{\nu}(xy)\,(xy)^{1/2}\,\dd x
\]
has the following inverse transform 
\[
f(x)=\frac{1}{\pi i}\int_{c-i\infty}^{c+i\infty}\hat{f}(y)I_{\nu}(xy)\,(xy)^{1/2}\,\dd y
\]
where $c$ is an arbitrary real number and $I_{\nu}(x)$ is the modified
Bessel function of the first kind. This is developed in a series of
papers \cite{meijer1940a,meijer1940b,Boas1942a,Boas1942b} and summarized
in Chapter X of Ref.~\cite{bateman1954tables}, which we simply quote
without going into further details of the proof. This gives us a unitary
transformation for the radial component, from which we obtain the
full 4D forward and inverse map: given a scalar function $\Psi(\eta,\rho,z,\bar{z})$,
we define the following analogue of the Fourier transform on four-dimensional
Minkowski spacetime 
\[
\hat{\Psi}(p,M,l,m)=\int_{-\infty}^{+\infty}d\eta\,\bar{\phi}_{pl}(\eta)\cosh^{2}\eta\,\int_{0}^{\infty}d\rho K_{i\sqrt{p^{2}-1}}(M\rho)(M\rho)^{1/2}\,\int_{\mathbb{C}}\frac{4\dd z\dd\bar{z}}{(1+|z|^{2})^{2}}\bar{Y}_{lm}(z,\bar{z})\,\Psi(\eta,\rho,z,\bar{z})
\]
that maps any scalar function $\Psi(\eta,\rho,z,\bar{z})$ to ``Fourier''
coefficients $\hat{\Psi}$ labeled by $(p,M,l,m)$, where $p>1$,
$M>0$, $l\ge0$, and $-l\le m\le l$. We note this $\rho$ integral
is indeed convergent. This transformation is invertible, with the
inverse map given by 
\begin{align*}
\Psi(\eta,\rho,z,\bar{z})={}\frac{1}{\pi i} & \int_{1}^{+\infty}2p\tanh\left(\frac{\pi}{2}\sqrt{p^{2}-1}\right)\,\dd p\int_{-i\infty}^{+i\infty}I_{i\sqrt{p^{2}-1}}(M\rho)(M\rho)^{1/2}\,\dd M\\
 & \sum_{l=0}^{\infty}\sum_{m=-l}^{l}\,\phi_{pl}(\eta)\,Y_{lm}(z,\bar{z})\hat{\Psi}(p,M,l,m)
\end{align*}
Here $I_{\nu}(x)$ is the modified Bessel function of the first kind.

We are now in a position to apply the forward and inverse map described
above to each of the Weyl scalars of the gravitational modes computed
earlier. This allows us to identify two families of celestial sphere
operators that encode the 5 complex Weyl scalars for each variety
of perturbation (scalar and vector) which we can label as $\hat{\Psi}_{pMlm}^{S,\alpha}$
and $\hat{\Psi}_{pMlm}^{V,\alpha}$. Here $S,V$ refer to scalar and
vector, and $\alpha=0,\cdots,4$ labels the 5 Weyl scalars. The $l,m$
angular momentum space is conjugate to the 2-sphere coordinate space
$z,\bar{z}$ so in this sense we obtain a holographic mapping of the
gravitational modes to the celestial sphere.

These celestial sphere operators will automatically have the desired
conformal transformation properties as shown in \cite{Liu:2021tif}.
We therefore find that the procedure described above allows us to
build a celestial sphere description of the gravitational modes living
in the Minkowski bulk spacetime at leading order. We stress that this
simple procedure amounts to use kinematic information to organize
the modes in a convenient way. With interactions included, one can
still use this procedure to map bulk dynamics to the celestial sphere.
The important question then is whether the dynamics has any useful
description incorporating the conformal symmetry of the celestial
sphere as in the program advocated in \cite{Pasterski:2016qvg}. In
the case of AdS/CFT the analogous answer was the holographic theory
was simpler than the gravity theory, being a quantum field theory
with conformal symmetry. In the case of asymptotically flat spacetime
it remains unclear whether the celestial sphere theory is a quantum
field theory. It remains a logical possibility that the 4D gravitational
description will be the simplest way to describe dynamics of the theory.
\begin{acknowledgments}
C.L. and D.L. are supported in part by DOE grant de-sc0010010 Task
A. 
\end{acknowledgments}

\appendix

\section{Weyl Scalars for Scalar Perturbation}

The Weyl scalars are evaluated at linear order for the scalar perturbation
mode $h_{\mu\nu}^{\omega lm}$ defined above. Here we use $\phi'=\frac{\partial}{\partial r}\phi$.

\begin{align*}
\Psi_{0}= & -\frac{1}{4r^{3}\sqrt{\Gamma(l-m-1)}\sqrt{\Gamma(l-m)}\sqrt{\Gamma(l+m+1)}}\left(\text{\ensuremath{\phi}}\left(l^{2}+l-2r\omega(r\omega+i)\right)+2r(1-ir\omega)\phi'\right)\times\\
 & \left(\sqrt{\Gamma(l-m)}\left(\sqrt{\Gamma(l-m+1)}\sqrt{\Gamma(l+m+3)}e^{-i2\varphi}Y_{l}^{m+2}+\right.\right.\\
 & \left.(m-1)m\tan^{2}\left(\frac{\theta}{2}\right)\sqrt{\Gamma(l-m-1)}\sqrt{\Gamma(l+m+1)}Y_{l}^{m}\right)\\
 & \left.-2me^{-i\varphi}\tan\left(\frac{\theta}{2}\right)\sqrt{\Gamma(l-m-1)}\sqrt{\Gamma(l-m+1)}\sqrt{\Gamma(l+m+2)}Y_{l}^{m+1}\right)\\
\Psi_{1}= & \frac{1}{4\sqrt{2}r^{3}\sqrt{\Gamma(l-m)}\sqrt{\Gamma(l+m+1)}}e^{-i\varphi}\left(r\left(\left(l^{2}+l-2\right)\text{\ensuremath{\phi}}'+2r(1-ir\omega)\text{\ensuremath{\phi}}''\right)+\right.\\
 & \left.i\text{\ensuremath{\phi}}\left(\left(l^{2}+l+2\right)r\omega+2il(l+1)-2r^{3}\omega^{3}-2ir^{2}\omega^{2}\right)\right)\times\\
 & \left(\sqrt{\Gamma(l-m+1)}\sqrt{\Gamma(l+m+2)}Y_{l}^{m+1}-me^{i\varphi}\tan\left(\frac{\theta}{2}\right)\sqrt{\Gamma(l-m)}\sqrt{\Gamma(l+m+1)}Y_{l}^{m}\right)\\
\Psi_{2}= & -\frac{1}{8r^{3}}Y_{l}^{m}\left(r\left(r\left(l(l+1)\text{\ensuremath{\phi}}''+2r\phi'''\right)-2\left(l^{2}+l-r^{2}\omega^{2}\right)\text{\ensuremath{\phi}}'\right)+l(l+1)\phi\left(r^{2}\omega^{2}+2\right)\right)\\
\Psi_{3}= & \frac{1}{8\sqrt{2}r^{3}\sqrt{\Gamma(l-m)}\sqrt{\Gamma(l+m+1)}}e^{-i\varphi}\left(r\left(-\left(l^{2}+l-2\right)\text{\ensuremath{\phi}}'+2r(-1-ir\omega)\text{\ensuremath{\phi}}''\right)+\right.\\
 & \left.\text{\ensuremath{\phi}}\left(i\left(l^{2}+l+2\right)r\omega+2l(l+1)-2ir^{3}\omega^{3}-2r^{2}\omega^{2}\right)\right)\times\\
 & \left(\sqrt{\Gamma(l-m+1)}\sqrt{\Gamma(l+m+2)}Y_{l}^{m+1}+me^{i\varphi}\cot\left(\frac{\theta}{2}\right)\sqrt{\Gamma(l-m)}\sqrt{\Gamma(l+m+1)}Y_{l}^{m}\right)\\
\Psi_{4}= & -\frac{1}{16r^{3}\sqrt{\Gamma(l-m-1)}\sqrt{\Gamma(l-m)}\sqrt{\Gamma(l+m+1)}}e^{-i2\varphi}\left(\text{\ensuremath{\phi}}\left(l^{2}+l-2r\omega(r\omega-i)\right)+2r(1+ir\omega)\text{\ensuremath{\phi}}'\right)\times\\
 & \left(\sqrt{\Gamma(l-m)}\sqrt{\Gamma(l-m+1)}\sqrt{\Gamma(l+m+3)}Y_{l}^{m+2}+\right.\\
 & \left.me^{i\varphi}\cot\left(\frac{\theta}{2}\right)\sqrt{\Gamma(l-m-1)}\left(2\sqrt{\Gamma(l-m+1)}\sqrt{\Gamma(l+m+2)}Y_{l}^{m+1}+\right.\right.\\
 & \left.\left.(m-1)e^{i\varphi}\cot\left(\frac{\theta}{2}\right)\sqrt{\Gamma(l-m)}\sqrt{\Gamma(l+m+1)}Y_{l}^{m}\right)\right)\,.
\end{align*}

\section{Weyl Scalars for Vector Perturbation}

The Weyl scalars are evaluated at linear order for the vector perturbation
mode $h_{\mu\nu}^{\omega lm}$ defined above. Here we use $\phi'=\frac{\partial}{\partial r}\phi$.
For brevity of presentation we follow \cite{PhysRev.108.1063} and
present the $m=0$ modes. The $m\neq0$ modes can always be obtained
by acting with an $SO(3)$ generator.

\begin{align*}
\Psi_{0}= & \frac{\sqrt{\Gamma(l+3)}e^{-i2\varphi}Y_{l}^{2}\left(ir\text{\ensuremath{\phi}}''+2(r\omega+i)\text{\ensuremath{\phi}}'+\omega\text{\ensuremath{\phi}}(2-ir\omega)\right)}{2\sqrt{l(l+1)}r^{2}\sqrt{\Gamma(l-1)}}\\
\Psi_{1}= & -\frac{\sqrt{\Gamma(l+2)}e^{-i\varphi}Y_{l}^{1}\left(i\left(\left(r^{2}\omega^{2}-2\right)\phi'+r\left(r\phi'''+(2-ir\omega)\text{\ensuremath{\phi}}''(r)\right)\right)+\omega\phi(r\omega+(-1+i))(r\omega+(1+i))\right)}{2\sqrt{2}\sqrt{l(l+1)}r^{2}\sqrt{\Gamma(l)}}\\
\Psi_{2}= & -\frac{ie^{-i2\varphi}\left(r^{2}\phi''+\phi\left(r^{2}\omega^{2}-2\right)\right)\left(\sqrt{\Gamma(l)}\sqrt{\Gamma(l+3)}Y_{l}^{2}+2e^{i\varphi}\cot(\theta)\sqrt{\Gamma(l-1)}\sqrt{\Gamma(l+2)}Y_{l}^{1}\right)}{4\sqrt{l(l+1)}r^{3}\sqrt{\Gamma(l-1)}\sqrt{\Gamma(l)}}\\
\Psi_{3}= & \frac{\sqrt{\Gamma(l+2)}e^{-i\varphi}Y_{l}^{1}\left(i\left(\left(r^{2}\omega^{2}-2\right)\phi'+r\left(r\phi'''+(2+ir\omega)\text{\ensuremath{\phi}}''\right)\right)+w\phi(2+r\omega(-r\omega+2i))\right)}{4\sqrt{2}\sqrt{l(l+1)}r^{2}\sqrt{\Gamma(l)}}\\
\Psi_{4}= & \frac{\sqrt{\Gamma(l+3)}e^{-i2\varphi}Y_{l}^{2}\left(ir\phi''+(-2r\omega+2i)\text{\ensuremath{\phi}}'+\omega\text{\ensuremath{\phi}}(-2-ir\omega)\right)}{8\sqrt{l(l+1)}r^{2}\sqrt{\Gamma(l-1)}}\,.
\end{align*}

\bibliographystyle{utphys}
\bibliography{gravups}

\end{document}